\documentclass[12pt]{article}
\usepackage{graphicx}

\newcommand{\beq}{\begin{equation}}
\newcommand{\eeq}{\end{equation}}
\newcommand{\beqa}{\begin{eqnarray}}
\newcommand{\eeqa}{\end{eqnarray}}
\newcommand{\non}{\nonumber}
\newcommand{\lab}{\label}
\newcommand{\bra}{\langle}
\newcommand{\ket}{\rangle}

\begin{document}

\title{Interaction-free generation of entanglement}

\author{Hiroo Azuma\thanks{On leave from
Centre for Quantum Computation,
Clarendon Laboratory,
Parks Road, Oxford OX1 3PU, United Kingdom.}
\\
{\small Canon Research Center, 5-1,
Morinosato-Wakamiya,}\\
{\small Atsugi-shi, Kanagawa, 243-0193, Japan}\\
{\small E-mail: azuma.hiroo@canon.co.jp}}

\date{June 14, 2003}

\maketitle

\begin{abstract}
In this paper, we study how to generate entanglement
by interaction-free measurement.
Using Kwiat et al.'s interferometer,
we construct a two-qubit quantum gate
that changes a particle's trajectory
according to the other particle's trajectory.
We propose methods for generating the Bell state
from an electron and a positron
and from a pair of photons by this gate.
We also show that using this gate,
we can carry out the Bell measurement
with the probability of $3/4$ at the maximum
and execute a controlled-NOT operation
by the method proposed by Gottesman and Chuang
with the probability of $9/16$ at the maximum.
We estimate the success probability for generating
the Bell state by our procedure
under imperfect interaction.
\end{abstract}

\section{Introduction}
\lab{introduction}
As a lot of researchers study quantum information processing (QIP)
(for example, the quantum teleportation, quantum computational
algorithms, and so on), 
they recognize entanglement to be important
\cite{Bennett-Brassard,Bouwmeester,quantum-algorithms}.
At the same time, experimental methods to generate entanglement
are investigated.
The entanglement is a quantum mechanical correlation between two systems
that can be separated from each other locally.
For simplicity,
let us think only about pure states in quantum mechanics here.
If we cannot describe a whole state of systems $A$ and $B$
as a simple product of
$|\Psi_{AB}\ket=|\psi_{A}\ket\otimes|\phi_{B}\ket$,
we say that $|\Psi_{AB}\ket$ is entangled
or the systems $A$ and $B$ have entanglement.
When the two systems are entangled,
we cannot create them
by classical communication between $A$ and $B$
and arbitrary local operations to each of $A$ and $B$.
Therefore, an entangled state has a correlation
that cannot be explained
by the classical probability theory\cite{entanglement}.

Entanglement is considered to be an essential resource for QIP.
In QIP, we often consider qubits
that are two-state quantum systems.
The Bell states are typical entangled states,
each of which consists of two qubits.
We define $\{|0\ket,|1\ket\}$ as an orthonormal basis
of a two-dimensional Hilbert space where a qubit is defined.
The Bell states are given by
\beq
|\Phi^{\pm}\ket=(1/\sqrt{2})(|00\ket\pm|11\ket),
\quad\quad
|\Psi^{\pm}\ket=(1/\sqrt{2})(|01\ket\pm|10\ket).
\eeq
Because $\{|\Phi^{\pm}\ket,|\Psi^{\pm}\ket\}$ form an orthonormal basis
of a four-dimensional Hilbert space where two qubits are defined,
we call them the Bell basis.

The Bell state plays an important role in the quantum teleportation.
Besides, it is shown that if we can prepare a four-qubit entangled state,
\beq
|\chi\ket=(1/2)
[(|00\ket+|11\ket)|00\ket+(|01\ket+|10\ket)|11\ket],
\eeq
carry out the quantum teleportation
(measurement of all two qubits by the Bell basis),
and execute arbitrary one-qubit unitary transformations,
we can construct the controlled-NOT gate,
which is one of the fundamental gates\cite{Gottesman-Chuang}.
Hence, generating the Bell states and the Bell measurement
are milestones for QIP.

A pair of photons in the Bell state can be generated
by pumping ultraviolet pulses into a nonlinear crystal
(for example, $\mbox{BBO}$ (beta-barium borate) and
$\mbox{LiIO}_{3}$)\cite{Bouwmeester,Kwiat-Mattle}.
Applying the pulses, the parametric down-conversion is caused
and pairs of polarization-entangled photons are created.
Because a rate of the parametric down-conversion depends
on the second order of the nonlinear susceptibility $\chi^{(2)}$,
creation of Bell pairs is in a low rate
and we have to supply the pulses to the crystal
with strong intensity in the laboratory.
Although some experimental methods
of cavity quantum electrodynamics (cavity QED) are proposed
to construct a two-qubit gate
(for example, the controlled-NOT gate),
the realization of it is considered to be
in the distinct future
because of technical difficulties\cite{2-qubit-gate-cavityQED}.
The perfect Bell measurement is considered to be a hard goal
to achieve as well
because it requires a two-qubit gate in general.

In this paper, we study how to generate the Bell state
using interaction-free measurement (IFM)\cite{Elitzur-Vaidman}.
Kwiat et al.'s IFM changes a photon's trajectory according to
whether an absorptive object exists or not
in a interferometer\cite{Kwiat-Weinfurter}.
This implies that information of the absorptive object is
transmitted to the photon.
Hence, this is a kind of two-particle quantum gate operation.
Regarding the absorptive object as quantum rather than classical,
we construct the Bell state from the absorptive object
and the photon by Kwiat et al.'s interferometer.
We also study how to carry out the Bell measurement
and the controlled-NOT operation using IFM process.
We evaluate the success probability for generating
the Bell state by our method under noise.
(We estimate the success probability of IFM under imperfect interaction.)

This paper is organized as follows.
In section~\ref{GenerationBell-electron-positron},
we discuss how to generate the Bell state
from an electron and a positron
with Kwiat et al.'s interferometer.
In section~\ref{Generation-Bell-pair-photon},
we discuss how to generate the Bell state from a pair of photons.
In section~\ref{Bell-measurement-construction-controlled-NOT-gate},
we consider how to carry out the Bell measurement
and construct the controlled-NOT gate.
In section~\ref{interaction-free-process-imperfect-interaction},
we estimate the success probability for generating
the Bell state by our procedure
(the success probability of IFM)
under imperfect interaction.
In section~\ref{Discussion},
we give brief discussion.

In the rest of this section,
we give a short review of IFM.
IFM originates
from Elitzur and Vaidman's following problem.
`Let us assume there is an object that absorbs a photon
with strong interaction if the photon approaches the object
close enough.
Can we examine whether the object exists or not
without its absorption?'\cite{Elitzur-Vaidman}
The reason that we do not want to let the object absorb the photon is
that it might lead an explosion, for example.
Elitzur and Vaidman themselves present a method of IFM
that is inspired by the Mach-Zehnder interferometer.
Then, more refined one is proposed by Kwiat et al.\cite{Kwiat-Weinfurter}.
We introduce the latter here.
(Kwiat et al.'s method is implemented
with high efficiency in experiments\cite{Kwiat-White}.)

We consider an interferometer that consists of $N$ beam splitters
as Figure~\ref{KWinterferometer2}.
We describe the upper paths as $a$ and the lower paths as $b$,
so that beam splitters form the boundary line
between paths $a$ and paths $b$
in the interferometer.
The interferometer has two input ports
that are the upper left port of $a$
and the lower left port of $b$,
and two output ports that are upper right port of $a$
and the lower right port of $b$.
We describe a state that there is one photon on the paths $a$
as $|1\ket_{a}$,
and a state that there is not a photon on the paths $a$
as $|0\ket_{a}$.
We assume a relation of
${}_{a}\bra i|j\ket_{a}=\delta_{ij}$ $\forall i,j\in\{0,1\}$.
These things are applied to the paths of $b$ as well.
The beam splitter $B$ works as follows:
\beq
B:
\left\{
\begin{array}{rrr}
|1\ket_{a}|0\ket_{b} & \rightarrow &
\cos\theta|1\ket_{a}|0\ket_{b}-\sin\theta|0\ket_{a}|1\ket_{b}, \\
|0\ket_{a}|1\ket_{b} & \rightarrow &
\sin\theta|1\ket_{a}|0\ket_{b}+\cos\theta|0\ket_{a}|1\ket_{b}.
\end{array}
\right.
\lab{definition-beamsplitter-B}
\eeq

\begin{figure}
\begin{center}
\includegraphics[scale=0.9]{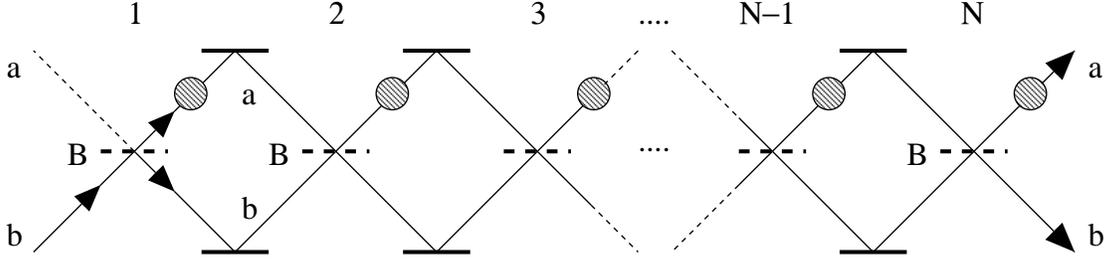}
\end{center}
\caption{Kwiat et al.'s interferometer for IFM.}
\lab{KWinterferometer2}
\end{figure}

Let us throw a photon into the lower left port of $b$.
If there is not the object on the paths,
a wave function of the photon that comes from the $k$th beam splitter
is given by
\beq
\sin k\theta|1\ket_{a}|0\ket_{b}
+\cos k\theta|0\ket_{a}|1\ket_{b}
\quad\quad
\mbox{for $k=0,1,...,N$.}
\eeq
Assuming $\theta=\pi/2N$,
the wave function of the photon that comes from the $N$th beam splitter
is equal to $|1\ket_{a}|0\ket_{b}$.
Hence, if there is not the object on the paths,
an incident photon from the lower left port of $b$ goes
to the upper right port of $a$.

Then, we consider the case where the object is on the paths of $a$
in the interferometer.
We assume that the object is put on every path of $a$
that comes from each beam splitter,
and all of these $N$ objects are the same one.
The photon thrown into the lower left port of $b$
cannot go to the upper right port of $a$
because the object absorbs it.
If the incident photon goes to the lower right port of $b$,
it has not passed through paths $a$ in the interferometer.
So that, the probability
that the photon goes to the lower right port of $b$
is equal to a product
of beam splitters' reflection rates,
and it is given by $P=\cos^{2N}\theta$.
In the large $N$ limit, we obtain
\beqa
\lim_{N\rightarrow\infty}P
&=&\lim_{N\rightarrow\infty}\cos^{2N}(\frac{\pi}{2N})
=\lim_{N\rightarrow\infty}
[1-\frac{\pi^{2}}{4N}+O(\frac{1}{N^{2}})]\non \\
&=&1.
\eeqa

Let us give a summary of the above discussion.
Kwiat et al.'s interferometer directs the incident photon
from the lower left port of $b$
with the probability of $P$ at least
as follows:
\begin{itemize}
\item if there is not the absorptive object in the interferometer,
the photon flies away to the upper right port of $a$,
\item if there is the absorptive object in the interferometer,
the photon flies away to the lower right port of $b$.
\end{itemize}
Furthermore, if we take large $N$, we can put $P$ arbitrarily close to one.

\section{Generation of the Bell state from an electron and a positron}
\lab{GenerationBell-electron-positron}
Kwiat et al.'s IFM directs an outgoing photon from the interferometer
to the ports of $a$ and $b$ in Figure~\ref{KWinterferometer2}
according to whether the absorptive object exists or not
on the paths $a$.
It admits an interpretation that information of the object is written
in the photon.
Moreover, because there is no annihilation of the photon
under the limit of $N\rightarrow\infty$,
the reduction of the state does not occur in Kwiat et al.'s IFM.
(That is, there is no dissipation and the system keeps coherence.)
Because the quantum state is never destroyed during the IFM process
under $N\rightarrow\infty$,
we can consider the object to be quantum rather than classical.
In section~\ref{introduction}, we have regarded the object as a classical one
that can take only one of two cases:
the case where the object exists
and the case where it does not exist.
However, in this section, we think that the object can take
a superposition of two orthogonal states
that the object exists and it does not exist,
and we throw it into the interferometer.
(A similar idea is mentioned by Kwiat et al.\cite{Kwiat-White}.
Hardy investigates the case where quantum object,
rather than classical one,
is thrown into Elitzur and Vaidman's IFM interferometer\cite{Hardy}.)

Because of the above considerations,
we can expect Kwiat et al.'s IFM interferometer
to work as a kind of a quantum gate.
Hence, we describe the interferometer of Figure~\ref{KWinterferometer2}
as a symbol of Figure~\ref{ifmgate2}.
In Figure~\ref{ifmgate2}, $x$ and $x'$ stand for paths of ports
that the absorptive object comes from and goes to.
$a$, $b$, $a'$, and $b'$ represent paths of ports for a photon.
$a$ and $a'$ stand for paths
of the upper left and the upper right ports
in Figure~\ref{KWinterferometer2},
and $b$ and $b'$ stand for paths of the lower left and the lower right ports
in Figure~\ref{KWinterferometer2}.
As we have explained in section~\ref{introduction},
we never throw the photon into the port of $a$ for the IFM process,
so that we draw $a$ as a dashed line and put a black rectangle
on the port of $a$ in the symbol of the gate in Figure~\ref{ifmgate2}.
From now on, we call the symbol of Figure~\ref{ifmgate2} the IFM gate.
Table~\ref{Table-truthtable-IFMgate} gives a truth table of the IFM gate
under the limit of $N\rightarrow\infty$.
In the table, a symbol of `1' represents
that there is the incident particle
(the object or the photon) on a path,
and a symbol of `0' represents that there is nothing on a path.
If the object is thrown into the path of $x$
as a superposition of `0' and `1',
the IFM gate works linearly
according to Table~\ref{Table-truthtable-IFMgate}.
(We pay attention to the following things.
If we throw a photon into the port of $a$ in Figure~\ref{ifmgate2},
it is absorbed and the whole process is not reversible.
Hence, strictly speaking,
the IFM gate is not unitary.)

\begin{figure}
\begin{center}
\includegraphics[scale=0.9]{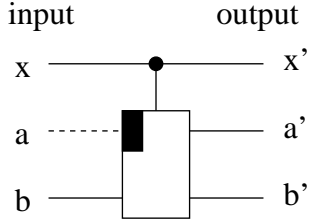}
\end{center}
\caption{The IFM gate that stands for Kwiat et al.'s interferometer.}
\lab{ifmgate2}
\end{figure}

\begin{table}
\begin{center}
\begin{tabular}{|ccc|ccc|}
\hline
\multicolumn{3}{|c|}{input} & \multicolumn{3}{c|}{output} \\
\hline
x & a & b & x' & a' & b' \\
\hline
0 & 0 & 1 & 0 & 1 & 0 \\
1 & 0 & 1 & 1 & 0 & 1 \\
\hline
\end{tabular}
\end{center}
\caption{A truth table of the IFM gate
under the limit of $N\rightarrow\infty$.}
\lab{Table-truthtable-IFMgate}
\end{table}

For simplicity, in the following discussion,
we replace the photon with an electron $e^{-}$
and replace the object with a positron $e^{+}$.
If the electron and the positron approach each other close enough,
they are annihilated together and a photon is created.
Here, we assume that this phenomenon occurs with the probability of one.
We can consider it to be that the positron absorbs the electron.
We can make beam splitters and mirrors for the electron and the positron
from plates that have suitable potential barriers.
Thus, we can construct the interferometer
of Figure~\ref{KWinterferometer2} for the electron and the positron
in the laboratory.

We pay attention to the following things.
To construct the IFM gate with an electron and a positron,
we have to adjust velocities and trajectories of them
so that they approach each other around circles of Figure~\ref{peKW-Bell1},
which represents Kwiat et al.'s interferometer for $e^{-}$ and $e^{+}$.
The electron and the positron considered here are quantum particles
and we have to regard them as wave packets that have fluctuations of
$\Delta \mbox{\boldmath $x$}$ and $\Delta \mbox{\boldmath $p$}$.
$|\Delta \mbox{\boldmath $x$}|$ is a typical size of the particle.
Because of uncertainty,
$|\Delta \mbox{\boldmath $x$}||\Delta \mbox{\boldmath $p$}|\sim \hbar$
is satisfied.
Let us suppose that a distance between the electron and the positron
has to be less than $\Delta r$ at time $t$
for their annihilation.
(We suppose that $\Delta r$ is a characteristic length
of the interaction.)
In our discussion, we assume
$|\Delta \mbox{\boldmath $x$}| \ll \Delta r$.
Hence, when we think about two particles' access to each other,
we do not need to consider quantum fluctuation of their wave packets.

\begin{figure}
\begin{center}
\includegraphics[scale=0.9]{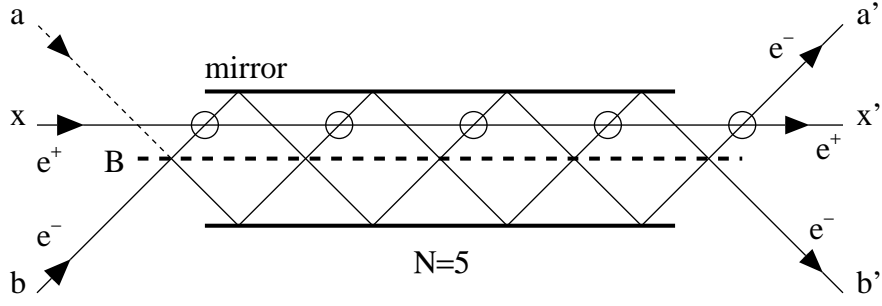}
\end{center}
\caption{Kwiat et al.'s interferometer
for an electron and a positron.}
\lab{peKW-Bell1}
\end{figure}

Let us consider a network in Figure~\ref{peBellcircuit3}.
Because it is a combination of quantum gates
as shown in Figure~\ref{peBellcircuit3},
we call it a quantum circuit.
$H$ stands for a beam splitter and it works as follows:
\beq
H:
\left\{
\begin{array}{rrr}
|0\ket_{x}|1\ket_{y} & \rightarrow &
(1/\sqrt{2})(|0\ket_{x}|1\ket_{y}+|1\ket_{x}|0\ket_{y}), \\
|1\ket_{x}|0\ket_{y} & \rightarrow &
(1/\sqrt{2})(|0\ket_{x}|1\ket_{y}-|1\ket_{x}|0\ket_{y}).
\end{array}
\right.
\lab{Hadamard-transformation}
\eeq
$H$ is called the Hadamard transformation.
The beam splitter $H$ transforms an incident positron from the path $y$
to a superposition of two paths with amplitude of $1/\sqrt{2}$ each.

\begin{figure}
\begin{center}
\includegraphics[scale=0.9]{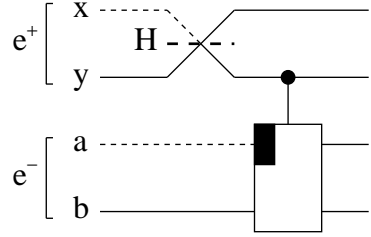}
\end{center}
\caption{A quantum circuit for generating the Bell state
of an electron and a positron.}
\lab{peBellcircuit3}
\end{figure}

We throw the positron into the path $y$
and the electron into the path $b$ as an initial state.
In Figure~\ref{peBellcircuit3},
time proceeds from the left side
to the right side and the whole state varies as follows:
\beqa
&& |0\ket_{x}|1\ket_{y}|0\ket_{a}|1\ket_{b} \non \\
& \stackrel{H}{\longrightarrow} &
(1/\sqrt{2})
(|0\ket_{x}|1\ket_{y}+|1\ket_{x}|0\ket_{y})|0\ket_{a}|1\ket_{b} \non \\
& \stackrel{\mbox{\scriptsize The IFM gate}}{\longrightarrow} &
(1/\sqrt{2})
(|0\ket_{x}|1\ket_{y}|0\ket_{a}|1\ket_{b}
+|1\ket_{x}|0\ket_{y}|1\ket_{a}|0\ket_{b}).
\lab{Bell-generation-transformation}
\eeqa
We define logical ket vectors of the electron and the positron
as follows:
\beq
\left\{
\begin{array}{rrr}
|\bar{0}\ket_{+} &=& |0\ket_{x}|1\ket_{y} \\
|\bar{1}\ket_{+} &=& |1\ket_{x}|0\ket_{y}
\end{array}
\right.,
\quad\quad
\left\{
\begin{array}{rrr}
|\bar{0}\ket_{-} &=& |0\ket_{a}|1\ket_{b} \\
|\bar{1}\ket_{-} &=& |1\ket_{a}|0\ket_{b}
\end{array}
\right.,
\lab{definition-logical-ket-vectors}
\eeq
where
${}_{\alpha}\bra\bar{i}|\bar{j}\ket_{\beta}
=\delta_{\alpha\beta}\delta_{ij}$
$\forall \alpha,\beta\in\{+,-\}$
$\forall i,j\in\{0,1\}$.
Then, we can rewrite the transformation of
Eq.~(\ref{Bell-generation-transformation})
as
\beq
|\bar{0}\ket_{+}|\bar{0}\ket_{-}
\rightarrow
|\Phi^{+}\ket
=(1/\sqrt{2})
(|\bar{0}\ket_{+}|\bar{0}\ket_{-}+|\bar{1}\ket_{+}|\bar{1}\ket_{-}).
\lab{Bell-generation-transformation-logicalket}
\eeq
This implies that we make the Bell state from a product state.

As shown in Eq.~(\ref{definition-logical-ket-vectors}),
we construct the logical ket vectors $\{|\bar{0}\ket,|\bar{1}\ket\}$
of a qubit from two paths,
so that this method is called
a dual-rail representation\cite{Chuang-Yamamoto}.
In this representation, only one particle must be always
on either of two paths.
Besides, putting a beam splitter
between two paths such as $B$ of Figure~\ref{KWinterferometer2}
and $H$ of Figure~\ref{peBellcircuit3},
we can carry out an arbitrary unitary transformation
on a two-dimensional Hilbert space that is generated
by $\{|\bar{0}\ket,|\bar{1}\ket\}$.
Applying a beam splitter to each pair of paths
that represents each qubit,
we can generate $|\Phi^{-}\ket$ and $|\Psi^{\pm}\ket$
from $|\Phi^{+}\ket$ generated
in Eq.~(\ref{Bell-generation-transformation-logicalket}).

In the quantum circuit of Figure~\ref{peBellcircuit3},
the number of electrons and that of positrons are preserved
during whole process,
so that pair annihilation never occurs.
Hence, generation of the Bell state
in Eqs.~(\ref{Bell-generation-transformation})
and (\ref{Bell-generation-transformation-logicalket})
is interaction-free process.

In a similar way, we can construct
Greenberger-Horne-Zeilinger (GHZ) state \\
$(1/\sqrt{2})(|\bar{0}\bar{0}\bar{0}\ket
+|\bar{1}\bar{1}\bar{1}\ket)$.
A quantum circuit shown in Figure~\ref{peGHZcircuit4}
causes the following transformation,
\beq
|\bar{0}\ket_{+}|\bar{0}\ket_{-}|\bar{0}\ket_{+}
\rightarrow
|\Phi^{+}\ket
=(1/\sqrt{2})
(|\bar{0}\ket_{+}|\bar{0}\ket_{-}|\bar{0}\ket_{+}
+|\bar{1}\ket_{+}|\bar{1}\ket_{-}|\bar{1}\ket_{+}).
\eeq

\begin{figure}
\begin{center}
\includegraphics[scale=0.9]{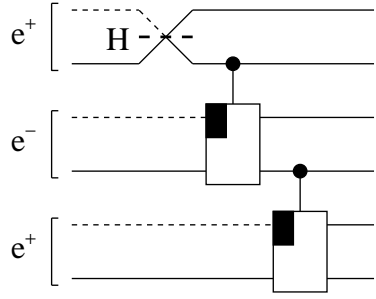}
\end{center}
\caption{A quantum circuit for generating the GHZ state.}
\lab{peGHZcircuit4}
\end{figure}

If we can create a positron with an accelerator,
we can construct the quantum circuit shown
in Figure~\ref{peBellcircuit3}
in a vacuum chamber.

\section{Generation of Bell state from a pair of photons}
\lab{Generation-Bell-pair-photon}
Let us consider how to create the Bell state from a pair of photons,
which are not annihilated by their accession to each other,
rather than an electron and a positron.
In this case, we need an atom that absorbs the photons.
For a while, we use only the Rabi oscillation
and beam splitters for photons as apparatus
because they are popular for cavity QED experiments.

We consider an atom that has three energy levels
as shown in Figure~\ref{atomlevel2}:
the ground state $g_{0}$,
the first excited state $e_{1}$,
and the second excited state $e_{2}$.
We assume that a difference of energy between $e_{1}$ and $g_{0}$
is equal to $\hbar\omega_{1}$,
and a difference of energy between $e_{2}$ and $g_{0}$
is equal to $\hbar\omega_{2}$.
We also assume that $\omega_{2}>\omega_{1}$,
and $\hbar\omega_{1}$, $\hbar\omega_{2}$, and $\hbar(\omega_{2}-\omega_{1})$
are large enough.
Besides, we assume that a transition between $e_{1}$ and $e_{2}$
is forbidden by some selection rule
although transitions between $g_{0}$ and $e_{1}$
and between $g_{0}$ and $e_{2}$ are admitted.
Defining a life time of spontaneous emission
$e_{1}\rightarrow g_{0}+\hbar\omega_{1}$ as $\tau_{1}$
and a life time of $e_{2}\rightarrow g_{0}+\hbar\omega_{2}$ as $\tau_{2}$,
we assume $\tau_{1}\gg\tau_{2}$.

\begin{figure}
\begin{center}
\includegraphics[scale=0.9]{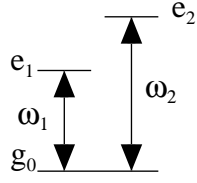}
\end{center}
\caption{Energy levels of the atom for the IFM gate.}
\lab{atomlevel2}
\end{figure}

If we supply an electric field (a laser pulse)
detuned slightly from the frequency $\omega_{1}$ to the atom,
it causes the Rabi oscillation between the ground state $g_{0}$
and the first excited state $e_{1}$\cite{Loudon}.
(The frequency of the pulse is given by
$\omega=\omega_{1}-\Delta\omega$, where
$0<|\Delta\omega|\ll\omega_{1}$.)
We regard a photon whose frequency is equal to $\omega_{2}$
as a qubit.
The atom of $g_{0}$ can absorb the photon $\omega_{2}$,
and the atom of $e_{1}$ cannot absorb the photon $\omega_{2}$.
We use this fact for the IFM process.
We assume that if we apply the photon $\omega_{2}$
to the atom of $g_{0}$,
it is absorbed with the probability of one.

Let us consider a quantum circuit
shown in Figure~\ref{photonBellcircuit2}.
We throw the atom of $|g_{0}\ket$ into the path of $x$
and the photons of $\omega_{2}$ into paths of $b$ and $d$.
We rewrite states of the atom on the path $x$ as
\beq
|0\ket_{x}=|e_{1}\ket,
\quad\quad
|1\ket_{x}=|g_{0}\ket.
\eeq
Applying a combination of some laser pulses
to the atom and causing the Rabi oscillations,
we can construct the Hadamard transformation $H'$ given by
\beq
H':
\left\{
\begin{array}{rrr}
|0\ket_{x} & \rightarrow &
(1/\sqrt{2})(|0\ket_{x}+|1\ket_{x}) \\
|1\ket_{x} & \rightarrow &
(1/\sqrt{2})(|0\ket_{x}-|1\ket_{x})
\end{array}
\right..
\eeq
In the quantum circuit of Figure~\ref{photonBellcircuit2},
the whole state varies as follows:
\beqa
&& |0\ket_{x}|0\ket_{a}|1\ket_{b}|0\ket_{c}|1\ket_{d} \non \\
& \stackrel{H'}{\longrightarrow} &
(1/\sqrt{2})
(|0\ket_{x}+|1\ket_{x})|0\ket_{a}|1\ket_{b}|0\ket_{c}|1\ket_{d} \non \\
& \stackrel{\mbox{\scriptsize The 1st IFM gate}}{\longrightarrow} &
(1/\sqrt{2})
(|0\ket_{x}|1\ket_{a}|0\ket_{b}+|1\ket_{x}|0\ket_{a}|1\ket_{b})
|0\ket_{c}|1\ket_{d} \non \\
& \stackrel{\mbox{\scriptsize The 2nd IFM gate}}{\longrightarrow} &
(1/\sqrt{2})
(|0\ket_{x}|1\ket_{a}|0\ket_{b}|1\ket_{c}|0\ket_{d}
+|1\ket_{x}|0\ket_{a}|1\ket_{b}|0\ket_{c}|1\ket_{d}) \non \\
&&\quad=
(1/\sqrt{2})
(|0\ket_{x}|\bar{1}\ket|\bar{1}\ket
+|1\ket_{x}|\bar{0}\ket|\bar{0}\ket) \non \\
& \stackrel{H'}{\longrightarrow} &
(1/2)
[|0\ket_{x}(|\bar{0}\ket|\bar{0}\ket+|\bar{1}\ket|\bar{1}\ket)
-|1\ket_{x}(|\bar{0}\ket|\bar{0}\ket-|\bar{1}\ket|\bar{1}\ket)].
\eeqa

\begin{figure}
\begin{center}
\includegraphics[scale=0.9]{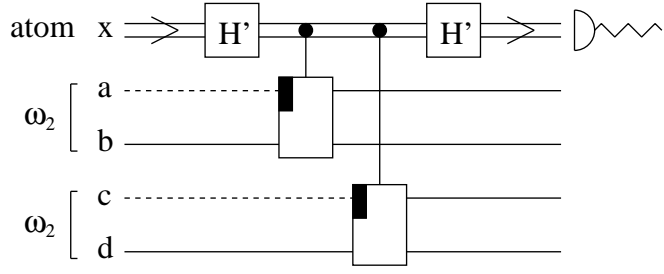}
\end{center}
\caption{A quantum circuit for generating
the Bell state from a pair of photons.}
\lab{photonBellcircuit2}
\end{figure}

Finally, we measure the atom by the orthonormal basis
$\{|0\ket_{x},|1\ket_{x}\}=\{|g_{0}\ket,|e_{1}\ket\}$.
If we obtain $|0\ket_{x}=|e_{1}\ket$,
the two photons are projected to the state
$|\Phi^{+}\ket
=(1/\sqrt{2})(|\bar{0}\bar{0}\ket+|\bar{1}\bar{1}\ket)$.
On the other hand, if we obtain $|1\ket_{x}=|g_{0}\ket$,
they are projected to the state
$|\Phi^{-}\ket
=(1/\sqrt{2})(|\bar{0}\bar{0}\ket-|\bar{1}\bar{1}\ket)$.
Hence, we obtain the Bell state of two photons.

Some methods that use technique of cavity QED are proposed
for generating entanglement
between two systems\cite{2-qubit-gate-cavityQED}.
Our method is essentially different from them.

\section{The Bell measurement and construction of the\\
controlled-NOT gate}
\lab{Bell-measurement-construction-controlled-NOT-gate}
Let us consider how to distinguish two-qubit states
$\{|\Phi^{\pm}\ket,|\Psi^{\pm}\ket\}$
that consist of an electron and a positron with the IFM gate.
We put one of $\{|\Phi^{\pm}\ket,|\Psi^{\pm}\ket\}$
into a quantum circuit shown in Figure~\ref{Bellmeasurement2}.
Although we have not thrown any particle into the path $a$
of the IFM gate in the former sections,
we consider the case where an electron is thrown into the path $a$
in this section.
If we throw a positron into the path $x$
and throw an electron into the path $a$,
pair annihilation of them occurs and a photon $\gamma$ is created
($e^{+}e^{-}\rightarrow\gamma$).
In this case,
the system suffers dissipation (or decoherence)
and it cannot work as a reversible quantum gate.

\begin{figure}
\begin{center}
\includegraphics[scale=0.9]{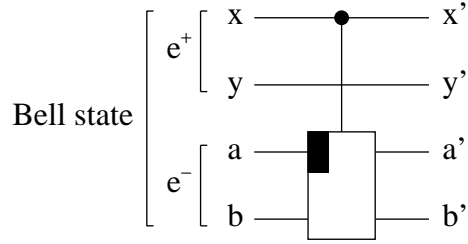}
\end{center}
\caption{A quantum circuit for the Bell measurement.}
\lab{Bellmeasurement2}
\end{figure}

Table~\ref{Table-truthtable-Bell-measurement}
is a truth table of the quantum circuit of Figure~\ref{Bellmeasurement2}
for the inputs of two-qubit states
$\{
|\bar{0}\ket_{+}|\bar{0}\ket_{-},
|\bar{0}\ket_{+}|\bar{1}\ket_{-},
|\bar{1}\ket_{+}|\bar{0}\ket_{-},
|\bar{1}\ket_{+}|\bar{1}\ket_{-}\}$,
where
$|\bar{0}\ket_{+}=|0\ket_{x}|1\ket_{y}$,
$|\bar{1}\ket_{+}=|1\ket_{x}|0\ket_{y}$,
$|\bar{0}\ket_{-}=|0\ket_{a}|1\ket_{b}$,
and $|\bar{1}\ket_{-}=|1\ket_{a}|0\ket_{b}$.
(In Table~\ref{Table-truthtable-Bell-measurement},
we take the limit of $N\rightarrow\infty$,
where $N$ is the number of the beam splitters.)
A symbol of $\gamma$ in the fourth row
of Table~\ref{Table-truthtable-Bell-measurement}
represents the pair annihilation of the electron and the positron.
Because we assume that we put one of $\{|\Phi^{\pm}\ket,|\Psi^{\pm}\ket\}$
into the quantum circuit,
we obtain $|0\ket_{b}$ for $|\Phi^{\pm}\ket$
and $|1\ket_{b}$ for $|\Psi^{\pm}\ket$
from the measurement of the path $b'$.
(We remember that $|\Phi^{\pm}\ket$ are superpositions
of $|\bar{0}\ket_{+}|\bar{0}\ket_{-}$
and $|\bar{1}\ket_{+}|\bar{1}\ket_{-}$
and $|\Psi^{\pm}\ket$ are superpositions
of $|\bar{0}\ket_{+}|\bar{1}\ket_{-}$
and $|\bar{1}\ket_{+}|\bar{0}\ket_{-}$.)
Hence, we can distinguish $|\Phi^{\pm}\ket$ and $|\Psi^{\pm}\ket$
by a result of the measurement for the path $b'$
by the basis of $\{|0\ket_{b},|1\ket_{b}\}$.

\begin{table}
\begin{center}
\begin{tabular}{|cccc|cccc|}
\hline
\multicolumn{4}{|c|}{input} & \multicolumn{4}{c|}{output} \\
\hline
x & y & a & b & x' & y' & a' & b' \\
\hline
0 & 1 & 0 & 1 & 0        & 1 & 1 & 0 \\
0 & 1 & 1 & 0 & 0        & 1 & 0 & 1 \\
1 & 0 & 0 & 1 & 1        & 0 & 0 & 1 \\
1 & 0 & 1 & 0 & $\gamma$ & 0 & 0 & 0 \\
\hline
\end{tabular}
\end{center}
\caption{A truth table of the quantum circuit shown
in Figure~\ref{Bellmeasurement2}
under the limit of $N\rightarrow\infty$.}
\lab{Table-truthtable-Bell-measurement}
\end{table}

Then, we consider the case where $|\Psi^{\pm}\ket$
is thrown into the circuit of Figure~\ref{Bellmeasurement2}.
If we obtain $|1\ket_{b}$ as a result of measurement for the path $b'$,
the state has varied through the circuit as follows:
\beqa
&&|\Psi^{\pm}\ket
=(1/\sqrt{2})
(|\bar{0}\ket_{+}|\bar{1}\ket_{-}\pm|\bar{1}\ket_{+}|\bar{0}\ket_{-}) \non \\
&&\quad=(1/\sqrt{2})
(|0\ket_{x}|1\ket_{y}|1\ket_{a}|0\ket_{b}
\pm|1\ket_{x}|0\ket_{y}|0\ket_{a}|1\ket_{b}) \non \\
& \stackrel{\mbox{\scriptsize The IFM gate}}{\longrightarrow} &
(1/\sqrt{2})
(|0\ket_{x}|1\ket_{y}\pm|1\ket_{x}|0\ket_{y})
|0\ket_{a}|1\ket_{b} \non \\
& \stackrel{\mbox{\scriptsize The measurement of $b'$}}{\longrightarrow} &
(1/\sqrt{2})
(|0\ket_{x}|1\ket_{y}\pm|1\ket_{x}|0\ket_{y})|0\ket_{a} \non \\
&&\quad
=(1/\sqrt{2})
(|\bar{0}\ket_{+}\pm|\bar{1}\ket_{+})|0\ket_{a}.
\eeqa
Then, we apply the beam splitter of $H$
defined in Eq.~(\ref{Hadamard-transformation})
to the positron's paths of $x$ and $y$.
A function of $H$ in the logical ket vectors is given by
\beq
H:
\left\{
\begin{array}{rrr}
|\bar{0}\ket & \rightarrow & (1/\sqrt{2})(|\bar{0}\ket+|\bar{1}\ket) \\
|\bar{1}\ket & \rightarrow & (1/\sqrt{2})(|\bar{0}\ket-|\bar{1}\ket)
\end{array}
\right..
\lab{Hadamard-transformation-in-logical-ket-vector}
\eeq
Hence, the incident $|\Psi^{\pm}\ket$
is transformed as
\beq
\left\{
\begin{array}{rrr}
|\Psi^{+}\ket & \rightarrow & |\bar{0}\ket_{+}|0\ket_{a} \\
|\Psi^{-}\ket & \rightarrow & |\bar{1}\ket_{+}|0\ket_{a}
\end{array}
\right..
\eeq
Therefore, we can distinguish $\{|\Psi^{+}\ket,|\Psi^{-}\ket\}$
by the measurement of the paths $x$ and $y$
with the basis of  $\{|\bar{0}\ket_{+},|\bar{1}\ket_{+}\}$.

If $|\Phi^{\pm}\ket$ are thrown into the quantum circuit,
dissipation (or decoherence) is caused in the system
by annihilation of the electron and the positron,
and we cannot apply quantum operations any more to the system.
Hence, if we obtain $|0\ket_{b}$ for the measurement of the path $b'$,
we decide whether $|\Phi^{+}\ket$ or $|\Phi^{-}\ket$
at random by tossing a coin.

In summary,
we can distinguish $|\Psi^{+}\ket$ and $|\Psi^{-}\ket$ perfectly,
and we can distinguish $|\Phi^{+}\ket$ and $|\Phi^{-}\ket$
with the probability of $1/2$.
In the quantum teleportation,
we need to distinguish the four basis vectors
$\{|\Phi^{\pm}\ket,|\Psi^{\pm}\ket\}$
in the following state,
\beqa
|\psi\ket\otimes|\Phi^{+}\ket
&=&(1/2)
[ |\Phi^{+}\ket\otimes          |\psi\ket
+ |\Phi^{-}\ket\otimes\sigma_{z}|\psi\ket \non \\
&&\quad\quad
+ |\Psi^{+}\ket\otimes\sigma_{x}|\psi\ket
-i|\Psi^{-}\ket\otimes\sigma_{y}|\psi\ket],
\eeqa
where $|\psi\ket$ is an arbitrary one-qubit state.
The above equation is a superposition of the four Bell states
with equal probabilistic amplitude.
Hence, if we carry out the Bell measurement with the IFM gate
in the way we discussed above,
we can execute the quantum teleportation
with the probability of $3/4$ at the maximum. 

Next, we consider how to carry out the Bell measurement
to an arbitrary two-qubit state,
\beq
|\Sigma\ket
=c_{00}|\Phi^{+}\ket+c_{01}|\Phi^{-}\ket
+c_{10}|\Psi^{+}\ket+c_{11}|\Psi^{-}\ket,
\eeq
where $c_{ij}\in \mbox{\boldmath $C$}$(complex)
$\forall i,j\in\{0,1\}$,
$\sum_{i,j\in\{0,1\}}|c_{ij}|^{2}=1$.
We can distinguish $|\Psi^{\pm}\ket$ perfectly
and we can distinguish $|\Phi^{\pm}\ket$ with the probability of $1/2$
by the method explained above.
Hence, if we permute $\{|\Phi^{\pm}\ket,|\Psi^{\pm}\ket\}$ at random
and measure them,
we can distinguish the Bell basis vectors
with the probability of $3/4$ at the maximum on the average.

Let us consider the following permutation of the Bell basis.
We define one-qubit rotational operators
around $x$, $y$, and $z$ axes as
\beq
R_{k}(\theta)=\exp[-i(\theta/2)\sigma_{k}],
\quad
k\in\{x,y,z\},
\quad
0\leq\theta<4\pi,
\eeq
where 
$\sigma_{k}$ is the Pauli matrix given by
\beq
\sigma_{x}= 
\left(
\begin{array}{cc}
0 & 1 \\
1 & 0
\end{array}
\right),
\quad
\sigma_{y}= 
\left(
\begin{array}{cc}
0 & -i \\
i & 0
\end{array}
\right),
\quad
\sigma_{z}= 
\left(
\begin{array}{cc}
1 & 0 \\
0 & -1
\end{array}
\right).
\eeq
The following relations are satisfied,
\beq
R_{k}(\pi/2)
=(1/\sqrt{2})(I-i\sigma_{k}),
\quad
R_{k}(\pi)=-i\sigma_{k},
\eeq
where
\beq
I=\left(
\begin{array}{cc}
1 & 0 \\
0 & 1
\end{array}
\right),
\eeq
and $k\in\{x,y,z\}$.
Because of
\beq
[R_{x}(\pi/2)\otimes R_{x}(\pi/2)]
[R_{y}(\pi)\otimes I]
|\Psi^{+}\ket
=-|\Phi^{-}\ket,
\eeq
we obtain the following relation,
\beq
[R_{x}(\pi/2)\otimes R_{x}(\pi/2)]
[R_{y}(\pi)\otimes I]:
\left\{
\begin{array}{rrr}
|\Phi^{+}\ket & \rightarrow & -|\Psi^{-}\ket \\
|\Phi^{-}\ket & \rightarrow & -i|\Phi^{+}\ket \\
|\Psi^{+}\ket & \rightarrow & -|\Phi^{-}\ket \\
|\Psi^{-}\ket & \rightarrow & -i|\Psi^{+}\ket
\end{array}
\right..
\eeq

We show six permutation operators in Table~\ref{permutation-Bell-basis},
where we omit phase factors.
These six operations permute the two sets of vectors,
$\{\Phi^{\pm}\}$ and $\{\Psi^{\pm}\}$,
to arbitrary combinations.
We take one integer $k\in\{1,...,6\}$ at random
and apply the $k$th operator in Table~\ref{permutation-Bell-basis}
to $|\Sigma\ket$.
Because all of the operators in Table~\ref{permutation-Bell-basis}
can be constructed by combinations of one-qubit unitary transformations,
we can make them with beam splitters.
After these procedures, we carry out the Bell measurement
with the IFM gate.

\begin{table}
\begin{center}
\begin{tabular}{|c|c@{\vrule width 0.8pt}cccc|}
\hline
No. & operator & $\Phi^{+}$ & $\Phi^{-}$ & $\Psi^{+}$ & $\Psi^{-}$ \\
\noalign{\hrule height 0.8pt}
1 & $I\otimes I$ & $\Phi^{+}$ & $\Phi^{-}$ & $\Psi^{+}$ & $\Psi^{-}$ \\
2 & $A$          & $\Psi^{-}$ & $\Psi^{+}$ & $\Phi^{-}$ & $\Phi^{+}$ \\
3 & $B$          & $\Phi^{+}$ & $\Psi^{+}$ & $\Phi^{-}$ & $\Psi^{-}$ \\
4 & $C$          & $\Psi^{+}$ & $\Phi^{-}$ & $\Phi^{+}$ & $\Psi^{-}$ \\
5 & $BA$         & $\Psi^{-}$ & $\Phi^{-}$ & $\Psi^{+}$ & $\Phi^{+}$ \\
6 & $CA$         & $\Psi^{-}$ & $\Phi^{+}$ & $\Phi^{-}$ & $\Psi^{+}$ \\
\hline
\end{tabular}
\end{center}
\caption{Permutation operators of Bell basis, where
$A=R_{y}(\pi)\otimes I$,
$B=R_{y}(\pi/2)\otimes R_{y}(\pi/2)$,
and $C=R_{x}(\pi/2)\otimes R_{x}(\pi/2)$.}
\lab{permutation-Bell-basis}
\end{table}

Then, we consider how to construct the controlled-NOT gate
by the IFM process.
The IFM gate defined in Figure~\ref{ifmgate2} seems
to be a kind of controlled operation.
The path $x$ is a control part
and the paths $a$ and $b$ make up a target part.
Let us look at Table~\ref{Table-truthtable-IFMgate}.
If the control part is set to $|0\ket_{x}$,
the target part is flipped from $|0\ket_{a}|1\ket_{b}$
to $|1\ket_{a}|0\ket_{b}$.
If the control part is set to $|1\ket_{x}$,
the target part of $|0\ket_{a}|1\ket_{b}$ is unchanged.
However, we cannot throw a photon into the path $a$
when there is an absorptive object on the path $x$.
From this consideration, we notice that it is difficult
to construct the controlled-NOT gate
that can be applied to arbitrary input states
from the IFM gate directly.

Gottesman and Chuang show that the controlled-NOT gate can be constructed
by preparing a four-qubit entangled state,
\beq
|\chi\ket=(1/2)
[(|00\ket+|11\ket)|00\ket+(|01\ket+|10\ket)|11\ket],
\eeq
and executing the Bell measurement
(the quantum teleportation) twice
as shown in Figure~\ref{GCcnotcircuit1}\cite{Gottesman-Chuang}.
Using this method, we can construct the controlled-NOT gate
that can be applied to arbitrary input states
with a certain probability from the IFM gate.
Thus, we discuss how to construct $|\chi\ket$ by the IFM gate.

\begin{figure}
\begin{center}
\includegraphics[scale=0.9]{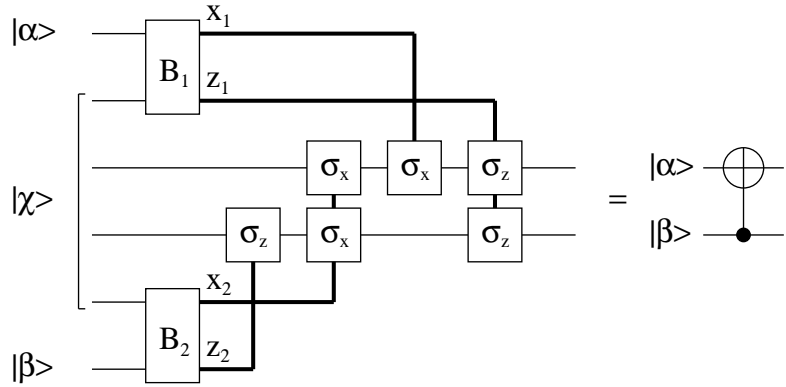}
\end{center}
\caption{A quantum circuit for the controlled-NOT
operation using the four-qubit entangled state
$|\chi\ket$\cite{Gottesman-Chuang}.
A thin line transmits a qubit,
and a thick line transmits a classical bit.
$B_{1}$ and $B_{2}$ stand for the Bell measurements.
Outputs of $B_{i}$ ($i\in\{1,2\}$) are given by
two classical bits:
$(x_{i},z_{i})=(0,0)$ for $|\Phi^{+}\ket$,
$(0,1)$ for $|\Phi^{-}\ket$,
$(1,0)$ for $|\Psi^{+}\ket$,
and $(1,1)$ for $|\Psi^{-}\ket$.
We apply $\sigma_{x}$ or $\sigma_{z}$
for $x_{i}=1$ or $z_{i}=1$,
and we do nothing for $x_{i}=0$ or $z_{i}=0$.}
\lab{GCcnotcircuit1}
\end{figure}

Making the GHZ state by the quantum circuit in Figure~\ref{peGHZcircuit4},
we put it into the upper three pairs of paths (three qubits)
of a quantum circuit shown in Figure~\ref{pe4qbtcircuit2}.
Then, we apply the Hadamard transformation
defined in Eq.~(\ref{Hadamard-transformation-in-logical-ket-vector})
to each of the three qubits
by a beam splitter as
\beqa
&&
(1/\sqrt{2})
(|\bar{0}\ket_{+}|\bar{0}\ket_{-}|\bar{0}\ket_{+}
+|\bar{1}\ket_{+}|\bar{1}\ket_{-}|\bar{1}\ket_{+}) \non \\
& \stackrel{H\otimes H\otimes H}{\longrightarrow} &
(1/\sqrt{2})
[(|\bar{0}\ket_{+}|\bar{0}\ket_{-}+|\bar{1}\ket_{+}|\bar{1}\ket_{-})
|\bar{0}\ket_{+}
+(|\bar{0}\ket_{+}|\bar{1}\ket_{-}+|\bar{1}\ket_{+}|\bar{0}\ket_{-})
|\bar{1}\ket_{+}].
\eeqa
Then, we add $|\bar{0}\ket_{-}$ as the forth qubit
and apply the IFM gate to the third and forth qubits
as follows:
\beqa
&&
(1/\sqrt{2})
[(|\bar{0}\ket_{+}|\bar{0}\ket_{-}+|\bar{1}\ket_{+}|\bar{1}\ket_{-})
|\bar{0}\ket_{+} \non \\
&&\quad
+(|\bar{0}\ket_{+}|\bar{1}\ket_{-}+|\bar{1}\ket_{+}|\bar{0}\ket_{-})
|\bar{1}\ket_{+}]|\bar{0}\ket_{-} \non \\
& \stackrel{\mbox{\scriptsize The IFM gate}}{\longrightarrow} &
(1/\sqrt{2})
[(|\bar{0}\ket_{+}|\bar{0}\ket_{-}+|\bar{1}\ket_{+}|\bar{1}\ket_{-})
|\bar{0}\ket_{+}|\bar{0}\ket_{-} \non \\
&&\quad
+(|\bar{0}\ket_{+}|\bar{1}\ket_{-}+|\bar{1}\ket_{+}|\bar{0}\ket_{-})
|\bar{1}\ket_{+}|\bar{1}\ket_{-}] \non \\
&&\quad=|\chi\ket.
\eeqa
(We use the fact that the IFM gate transforms
$|\bar{0}\ket_{+}|\bar{0}\ket_{-}
\rightarrow
|\bar{0}\ket_{+}|\bar{0}\ket_{-}$
and
$|\bar{1}\ket_{+}|\bar{0}\ket_{-}
\rightarrow
|\bar{1}\ket_{+}|\bar{1}\ket_{-}$
when $e^{+}$ is a control qubit and $e^{-}$ is a target qubit.
We can derive it from Table~\ref{Table-truthtable-IFMgate}.)

\begin{figure}
\begin{center}
\includegraphics[scale=0.9]{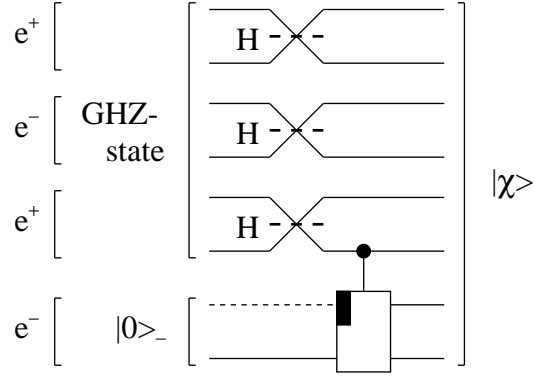}
\end{center}
\caption{A quantum circuit for generating the four-qubit entangled state
$|\chi\ket$.}
\lab{pe4qbtcircuit2}
\end{figure}

Therefore, if we can prepare the ideal IFM gate
under $N\rightarrow\infty$,
we can construct $|\chi\ket$ with the probability of one.
The success probability of the Bell measurement by our method
is equal to $3/4$
at the maximum,
and we have to carry out the Bell measurement twice
for the controlled-NOT gate
by Gottesman and Chuang's method.
Hence, the success probability of the controlled-NOT operation is given by
$(3/4)^{2}=4/9$ at the maximum.

\section{Interaction-free process under imperfect interaction}
\lab{interaction-free-process-imperfect-interaction}
The IFM gate discussed in the former sections are constructed
with beam splitters (one-qubit operations)
and interaction between a photon and an absorptive object.
In this section,
we consider the case where the interaction is not perfect.
(We regard the beam splitters as accurate enough.)
We assume that the photon is absorbed
with the probability of $(1-\eta)$
and it passes by the object without being absorbed
with the probability of $\eta$
when it approaches the object close.
We estimate the success probability of the IFM gate
under these assumptions.
It implies that we evaluate the success probability
for generating the Bell state by the IFM process.

We assume the following transformation
in Figure~\ref{KWinterferometer2}.
The photon that comes from each beam splitter to the upper path $a$ suffers
\beq
|\bar{1}\ket
\rightarrow
\sqrt{\eta}|\bar{1}\ket
+\sqrt{1-\eta}|\mbox{absorption}\ket,
\lab{imperfect-absorption-process}
\eeq
where $0<\eta<1$ and $|\bar{1}\ket=|1\ket_{a}|0\ket_{b}$.
$|\mbox{absorption}\ket$ is a state that the object absorbs the photon.
We assume that it is normalized and orthogonal to
$\{|\bar{0}\ket,|\bar{1}\ket\}$,
where
$|\bar{0}\ket=|0\ket_{a}|1\ket_{b}$
and
$|\bar{1}\ket=|1\ket_{a}|0\ket_{b}$.
(If we think about the interferometer with an electron and a positron,
it represents the state that a photon $\gamma$ is created
by pair annihilation of them.)

From now on, for simplicity,
we describe the transformations that are applied to the photon
as matrices by the basis of $\{|\bar{0}\ket,|\bar{1}\ket\}$.
Writing
\beq
|\bar{0}\ket=|0\ket_{a}|1\ket_{b}
=
\left(
\begin{array}{c}
1 \\
0
\end{array}
\right),
\quad
|\bar{1}\ket=|1\ket_{a}|0\ket_{b}
=
\left(
\begin{array}{c}
0 \\
1
\end{array}
\right),
\eeq
we can describe the beam splitter $B$ defined
in Eq.~(\ref{definition-beamsplitter-B}) as
\beq
B=
\left(
\begin{array}{cc}
\cos\theta & -\sin\theta \\
\sin\theta & \cos\theta
\end{array}
\right),
\lab{definition-matrix-B}
\eeq
where $\theta=\pi/2N$
and the absorption process defined
in Eq.~(\ref{imperfect-absorption-process}) as
\beq
A=
\left(
\begin{array}{cc}
1 & 0           \\
0 & \sqrt{\eta}
\end{array}
\right),
\lab{definition-matrix-A}
\eeq
where $0<\eta<1$.
The matrix $A$ is not unitary because the process
defined by Eq.~(\ref{imperfect-absorption-process})
causes the absorption of the photon (dissipation or decoherence).

The probability that an incident photon from the lower left port of $b$
passes through the $N$ beam splitters and
is detected in the lower right port of $b$ in Figure~\ref{KWinterferometer2}
is given by
\beq
P=|\bra\bar{0}|(BA)^{N-1}B|\bar{0}\ket|^{2}.
\lab{definition-fidelity-IFMgate}
\eeq
We plot results of numerical calculations of
the success probability $P$
as a function of $N$ and $\eta$ with
Eqs.~(\ref{definition-matrix-B}),
(\ref{definition-matrix-A}),
and (\ref{definition-fidelity-IFMgate})
and link them together by solid lines in Figure~\ref{KWifmrat2}.
In Figure~\ref{KWifmrat2},
four cases of $\eta=0$, $0.05$, $0.1$, and $0.2$
are shown in order from the top to the bottom.

\begin{figure}
\begin{center}
\includegraphics[scale=0.9]{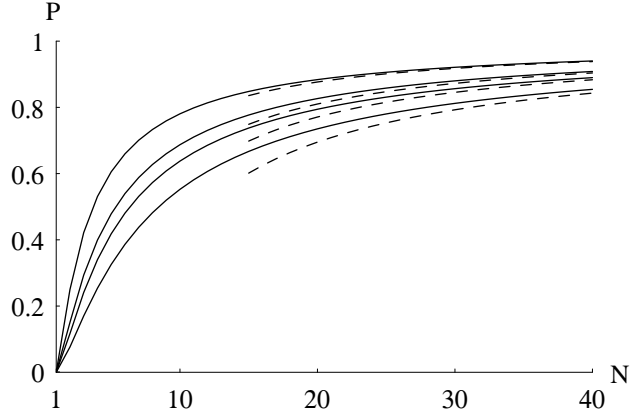}
\end{center}
\caption{Results of numerical calculation of
the success probability $P$
as a function of $N$ and $\eta$ linked with lines.
$\eta$ is the rate that the object fails in absorbing the photon.
$P$ and $\eta$ are dimensionless quantities.
$N$ is the number of the beam splitters.
Four cases of $\eta=0$, $0.05$, $0.1$, and $0.2$
are shown in order from the top to the bottom
as solid curves and dashed curves.
The solid curves represent exact results from
Eqs.~(\ref{definition-matrix-B}),
(\ref{definition-matrix-A}),
and (\ref{definition-fidelity-IFMgate}).
The dashed curves represent approximated results
from Eq.~(\ref{matrix-element-approximation}).}
\lab{KWifmrat2}
\end{figure}

We want to investigate how a variation of the noise $\eta$
affects the success probability $P$ of the IFM gate.
Fixing $\eta$ to some finite value,
we evaluate $P$ under the limit of $N\rightarrow\infty$.
When we try to estimate $P$ defined in Eq.~(\ref{definition-fidelity-IFMgate})
under $N\rightarrow\infty$,
we meet the following difficulty.
$P$ depends on $N$ from both
$\theta=\pi/2N$ of the matrix $B$ given by Eq.~(\ref{definition-matrix-B})
and
an exponent $N$ of $(BA)^{N-1}$ in Eq.~(\ref{definition-fidelity-IFMgate}).

We can treat the dependence of $\theta=\pi/2N$
with expanding $P$ by the order of $\theta$
and neglecting the higher order terms.
On the other hand,
we treat the exponent of $(BA)^{N-1}$ by the following way.
If we regard matrix elements of $BA$ as single terms,
each matrix element of $(BA)^{N-1}$ is given by
a sum of $2(N-2)$ terms.
Hence, estimating each matrix element of $BA$
up to the order of $\theta^{2}$
(that is, up to the order of $1/N^{2}$),
we can evaluate each matrix element of $(BA)^{N-1}$
up to the order of
$O(1/N^{2})\times 2(N-2)=O(1/N)$.

Let us estimate matrix elements up to the order of $\theta^{2}$
with fixing $\eta$ $(0<\eta<1)$
and taking $\theta=\pi/2N\rightarrow 0$.
Because we can expand $B$ as
\beq
B=
\left(
\begin{array}{cc}
1-(\theta^{2}/2) & -\theta          \\
\theta           & 1-(\theta^{2}/2)
\end{array}
\right)
+O(\theta^{3}),
\eeq
we obtain
\beqa
(BA)^{k}
&=&
\left(
\begin{array}{cc}
1-\theta^{2}[\sum_{l=1}^{k-1}l\sqrt{\eta}^{k-l}+(k/2)]
&
-\theta\sum_{l=0}^{k-1}\sqrt{\eta}^{l+1} \\
\theta\sum_{l=0}^{k-1}\sqrt{\eta}^{l} 
&
\sqrt{\eta}^{k}
-\theta^{2}[\sum_{l=1}^{k-1}l\sqrt{\eta}^{l}+(k/2)\sqrt{\eta}^{k}]
\end{array}
\right)
\non \\
&&\quad
+O(\theta^{3}),
\eeqa
by the induction, where $k=1, 2, ...$, and
$\sum_{l=1}^{0}$ means we do not take summation.
Hence, amplitude of the photon of the lower right port of $b$
that passes $N$ beam splitters
is given by
\beqa
&&
\bra\bar{0}|(BA)^{N-1}B|\bar{0}\ket \non \\
&=&
1-\theta^{2}(\frac{N}{2}+\sum_{l=0}^{N-2}\sqrt{\eta}^{l+1}
+\sum_{l=1}^{N-2}l\sqrt{\eta}^{N-1-l})
+O(\theta^{3}) \non \\
&=&
1-\theta^{2}(\frac{N}{2}+N\sum_{l=1}^{N-1}\sqrt{\eta}^{l}
-\sum_{l=1}^{N-1}l\sqrt{\eta}^{l})
+O(\theta^{3}) \non \\
&=&
1-(\frac{\pi}{2})^{2}(\frac{1}{N})
[\frac{1}{2}+\frac{\sqrt{\eta}(1-\sqrt{\eta}^{N-1})}{1-\sqrt{\eta}} \non \\
&&\quad
-\frac{1}{N}
\frac{\sqrt{\eta}[1-N\sqrt{\eta}^{N-1}+(N-1)\sqrt{\eta}^{N}]}
{(1-\sqrt{\eta})^{2}}]
+O(\frac{1}{N^{2}}),
\lab{matrix-element-approximation}
\eeqa
where $N=2, 3, ...$.
In the derivation of the above,
we use the following formulas,
\beq
\sum_{k=1}^{N}x^{k}=x(1-x^{N})/(1-x),
\eeq
\beq
\sum_{k=1}^{N}kx^{k}=x[1-(N+1)x^{N}+Nx^{N+1}]/(1-x)^{2}.
\eeq

Eq.~(\ref{matrix-element-approximation})
gives an approximation of
$\bra\bar{0}|(BA)^{N-1}B|\bar{0}\ket$
up to the order of $O(1/N)$ in the large $N$ limit.
We plot results of numerical estimation of
$P=|\bra\bar{0}|(BA)^{N-1}B|\bar{0}\ket|^{2}$
as a function of $N$ and $\eta$ up to the order of $O(1/N)$
with Eq.~(\ref{matrix-element-approximation})
and link them together by dashed lines in Figure~\ref{KWifmrat2}.
In Figure~\ref{KWifmrat2},
four cases of $\eta=0$, $0.05$, $0.1$, and $0.2$
are shown in order from the top to the bottom.
We can find that estimation of $P$
by Eq.~(\ref{matrix-element-approximation})
is a good approximation of exact numerical calculations
in the large $N$ limit.

Besides, we notice the following things
from Eq.~(\ref{matrix-element-approximation}).
Even if we fix $\eta$ to any value of $0<\forall\eta<1$,
we can put
$|\bra\bar{0}|(BA)^{N-1}B|\bar{0}\ket|^{2}$
arbitrarily close to one
by taking large $N$.
To put it in precise words,
if $0<\forall\eta<1$ and $0<\forall\epsilon<1$ are given,
we can take large $N_{0}$ so that
$[1-\bra\bar{0}|(BA)^{N-1}B|\bar{0}\ket]<\epsilon$
can be satisfied for all $N>N_{0}$.

This implies that
if there is no limitation on the number of beam splitters,
we can always overcome the noise
given by Eq.~(\ref{imperfect-absorption-process}).
If we let the transmission rate of the beam splitter be small enough
(take the large $N$ limit or take the small $\theta$ limit),
the probability that the photon approaches the absorptive object close
gets smaller.
Hence, we can suppress the contribution of the probability $\eta$
that the object fails in absorbing the photon.
If we take extremely large $N$,
the transmission rate of the beam splitter
$T=\sin^{2}(\pi/2N)$ takes a very small value.
This requires a beam splitter that has high precision
of $T=\sin^{2}(\pi/2N)\sim O(1/N^{2})$.
Hence, we can obtain an interpretation that the accuracy of the beam splitter
compensates the imperfect interaction
between the photon and the absorptive object.

We estimate how many beam splitters do we need
to let the success probability of the IFM gate
reach some given value $P$ $(0<\forall P<1)$
as an function of $\eta$.
If $N$ increases extremely as $\eta$ increases from zero,
we cannot say that the IFM gate is tolerant of imperfect interaction.
Let us assume that $\eta$ is comparatively small
and $N\sqrt{\eta}^{N}\ll1$ is satisfied for large $N$.
From Eq.~(\ref{matrix-element-approximation}),
the simple representation of $P$ is given by
\beq
P\sim
1-\frac{\pi^{2}}{2N}(\frac{1}{2}
+\frac{\sqrt{\eta}}{1-\sqrt{\eta}})
+O(\frac{1}{N^{2}}).
\eeq
We obtain
\beq
N\sim
(\frac{\pi}{2})^{2}
\frac{1}{1-P}
\frac{1+\sqrt{\eta}}{1-\sqrt{\eta}}.
\lab{estimation-number-beamsplitters}
\eeq
Eq.~(\ref{estimation-number-beamsplitters})
is valid when $P$ is close to one enough
(or $N$ is large enough).
In Eq.~(\ref{estimation-number-beamsplitters}),
$N$ gets larger extremely as $\eta$ increases
and
$\lim_{\eta\rightarrow 1}N=\infty$.

\section{Discussion}
\lab{Discussion}
In this paper,
we study how to generate the Bell state with interaction-free process.
As far as we use only a finite number of beam splitters,
the success probability of the IFM gate is lower than one.
Hence, our method for generating the Bell state is probabilistic
in the laboratory.
If we want to let the success probability of the gate get close to one,
we need to prepare a large number of beam splitters
and let the transmission rate of them get smaller.
Besides, accuracy of the beam splitters compensates the imperfect interaction.
Hence, although our method has simple mechanism as shown in this paper,
it requires a lot of highly accurate beam splitters.

In section~\ref{GenerationBell-electron-positron},
we discuss the quantum circuit that generates the Bell state
from an electron and a positron.
We may construct it in a semiconductor
using a hole and a conductive electron.

E.~Knill et al. propose schemes of non-deterministic
quantum gates with linear optical devices\cite{Knill-Laflamme}.
Motivation of their work is similar to ours.
P.~Horodecki investigates the dynamical evolution of the IFM process,
where the absorptive object is replaced
with the atom evolving coherently\cite{Horodecki}.

\bigskip
\noindent
{\bf \large Acknowledgements}
\smallskip

We thank M.~Okuda for helpful discussion
about section~\ref{Generation-Bell-pair-photon}.
We also thank L.~Vaidman and P.~Horodecki
for useful comments about references.


\begin{thebibliography}{99}
%
\bibitem{Bennett-Brassard}
C.~H.~Bennett, G.~Brassard, C.~Cr{\'e}peau,
R.~Jozsa, A.~Peres, and W.~K.~Wootters,
`Teleporting an unknown quantum state
via dual classical and Einstein-Podolsky-Rosen channels',
Phys. Rev. Lett. {\bf 70}, 1895--1899 (1993).
%
\bibitem{Bouwmeester}
D.~Bouwmeester, J.-W.~Pan, K.~Mattle,
M.~Eibl, H.~Weinfurter, and A.~Zeilinger,
`Experimental quantum teleportation',
Nature {\bf 390}, 575--579 (1997).
%
\bibitem{quantum-algorithms}
D.~Deutsch and R.~Jozsa,
`Rapid solution of problems by quantum computation',
Proc. R. Soc. Lond. A {\bf 439}, 553--558 (1992);\\
D.~R.~Simon,
`On the power of quantum computation',
SIAM J. Comput. {\bf 26}, 1474--1483 (1997);\\
P.~W.~Shor,
`Polynomial-time algorithms for prime factorization
and discrete logarithms on a quantum computer',
SIAM J. Comput. {\bf 26}, 1484--1509 (1997);\\
L.~K.~Grover,
`Quantum mechanics helps in searching for a needle in a haystack',
Phys. Rev. Lett. {\bf 79}, 325--328 (1997).
%
\bibitem{entanglement}
J.~S.~Bell,
{\it Speakable and unspeakable in quantum mechanics}
(Cambridge University Press,
Cambridge, 1987);\\
C.~H.~Bennett, D.~P.~DiVincenzo, J.~A.~Smolin, and W.~K.~Wootters,
`Mixed-state entanglement and quantum error correction',
Phys. Rev. A {\bf 54}, 3824--3851 (1996);\\
R.~F.~Werner,
`Quantum states with Einstein-Podolsky-Rosen correlations
admitting a hidden-variable model',
Phys. Rev. A {\bf 40}, 4277--4281 (1989);\\
S.~Popescu,
`Bell's inequalities and density matrices:
revealing ``hidden'' nonlocality',
Phys. Rev. Lett {\bf 74}, 2619--2622 (1995).
%
\bibitem{Gottesman-Chuang}
D.~Gottesman and I.~L.~Chuang,
`Demonstrating the viability of universal quantum computation
using teleportation and single-qubit operations',
Nature {\bf 402}, 390--393 (1999).
%
\bibitem{Kwiat-Mattle}
P.~G.~Kwiat, K.~Mattle, H.~Weinfurter, A.~Zeilinger, A.~V.~Sergienko,
and Y.~Shih,
`New high-intensity source of polarization-entangled photon pairs',
Phys. Rev. Lett {\bf 75}, 4337--4341 (1995).
%
\bibitem{2-qubit-gate-cavityQED}
Q.~A.~Turchette, C.~J.~Hood, W.~Lange, H.~Mabuchi, and H.~J.~Kimble,
`Measurement of conditional phase shifts for quantum logic',
Phys. Rev. Lett. {\bf 75}, 4710--4713 (1995);\\
C.~Monroe, D.~M.~Meekhof, B.~E.~King, W.~M.~Itano, and D.~J.~Wineland,
`Demonstration of a fundamental quantum logic gate',
Phys. Rev. Lett. {\bf 75}, 4714--4717 (1995).
%
\bibitem{Elitzur-Vaidman}
A.~C.~Elitzur and L.~Vaidman,
`Quantum mechanical interaction-free measurements',
Found. Phys. {\bf 23}, 987--997 (1993);\\
L.~Vaidman,
`Are interaction-free measurements interaction free?',
Opt. Spectrosc. {\bf 91}, 352--357 (2001).
%
\bibitem{Kwiat-Weinfurter}
P.~Kwiat, H.~Weinfurter, T.~Herzog, A.~Zeilinger, and M.~A.~Kasevich,
`Interaction-free measurement',
Phys. Rev. Lett. {\bf 74}, 4763--4766 (1995).
%
\bibitem{Kwiat-White}
P.~G.~Kwiat, A.~G.~White, J.~R.~Mitchell, O.~Nairz,
G.~Weihs, H.~Weinfurter, and A.~Zeilinger,
`High-efficiency quantum interrogation measurements
via the quantum Zeno effect',
Phys. Rev. Lett. {\bf 83}, 4725--4728 (1999).
%
\bibitem{Hardy}
L.~Hardy,
`Quantum mechanics, local realistic theories,
and Lorentz-invariant realistic theories',
Phys. Rev. Lett. {\bf 68}, 2981--2984 (1992).
%
\bibitem{Chuang-Yamamoto}
I.~L.~Chuang and Y.~Yamamoto,
`Simple quantum computer',
Phys. Rev. A {\bf 52}, 3489--3496 (1995).
%
\bibitem{Loudon}
R.~Loudon,
{\it The quantum theory of light, second edition}
(Oxford University Press,
Oxford, 1983), Chap.~2.
%
\bibitem{Knill-Laflamme}
E.~Knill, R.~Laflamme, and G.~J.~Milburn,
`A scheme for efficient quantum computation with linear optics',
Nature {\bf 409}, 46--52 (2001).
%
\bibitem{Horodecki}
P.~Horodecki,
` ``Interaction-free'' interaction:
Entangling evolution coming from the possibility of detection',
Phys. Rev. A {\bf 63}, 022108 (2001).
\end{thebibliography}
\end{document}